\documentclass[]{achemso}
\usepackage[version=3]{mhchem} 
\usepackage{amsmath}
\usepackage{graphicx}

\author{Ershad Mohammadi}
\affiliation[]
{Laboratory of Nanoscience for Energy Technologies (LNET), Faculty of
Engineering (STI), Ecole Polytechnique Federale de Lausanne (EPFL), Lausanne 1015, Switzerland}

\alsoaffiliation[Research]
{Department of Information in Matter and Center for Nanophotonics, AMOLF, Science Park 104, 1098 XG Amsterdam, Netherlands.}

\author{Giulia Tagliabue}
\email{giulia.tagliabue@epfl.ch}
\affiliation[]
{Laboratory of Nanoscience for Energy Technologies (LNET), Faculty of
Engineering (STI), Ecole Polytechnique Federale de Lausanne (EPFL), Lausanne 1015, Switzerland}

\title[An \textsf{achemso} demo]
  {Nanophotonic-Enhanced Thermal Circular Dichroism for Chiral Sensing}

\keywords{Chiral sensing, Thermal circular dichroism, Thermonanophotonics, Optical chirality, High-index dielectric metasurfaces, nonlocal metasurfaces}

\begin{document}

\begin{abstract}
Circular Dichroism (CD) can distinguish the handedness of chiral molecules. However, it is typically very weak due to vanishing absorption at low molecular concentrations. Here, we suggest Thermal Circular Dichroism (TCD) for chiral detection, leveraging the temperature difference in the chiral sample when subjected to right- and left-circularly polarized excitations. The TCD combines the enantiospecificity of circular dichroism with the higher sensitivity of thermal measurements, while introducing new opportunities in the thermal domain that can be synergistically combined with optical approaches.

We propose a theoretical framework to understand the TCD of individual and arrays of resonators covered by chiral molecules. To enhance the weak TCD of chiral samples, we first use individual dielectric Mie resonators and identify chirality transfer and self-heating as the underlying mechanisms giving rise to the differential temperature. However, inherent limitations imposed by the materials and geometries of such resonators make it challenging to surpass a certain level in enhancements. To overcome this, we suggest nonlocal thermal and electromagnetic interactions in arrays. We predict that a combination of chirality transfer to Mie resonators, collective thermal effects, and optical lattice resonance could, in principle, offer more than 4 orders of magnitude enhancement in TCD. Our thermonanophotonic-based approach thus establishes key concepts for ultrasensitive chiral detection.

\end{abstract}

Chiral molecules exist in right- and left-handed configurations, each with distinct properties. Chiral sensing is the differentiation between these two forms, which is crucial for gaining insight into molecular structure, drug development, and monitoring environmental pollutants, among many applications.\cite{barron1990chirality,kallenborn2001chiral,basheer2018chemical}. Circular Dichroism (CD) spectroscopy can discern this handedness by measuring the contrast in absorption of a chiral sample for right- and left-circularly polarized lights. The CD signal for a sample solution is quantified as $ \text{CD} = 32.98\Delta\epsilon\text{C}l$ (Supporting Information, Section S1), where $\Delta\epsilon$ is the differential molar attenuation of the chiral species inside the sample in  $\text{M}^{-1}\text{cm}^{-1}$, $\text{C}$ is the molar concentration in mol/L (usually denoted by M), and $l$ is the pathlength in $\text{cm}$.\cite{fasman2013circular,mohammadi2023nanophotonic} Given a typical value of $\Delta\epsilon = 20 \text{M}^{-1} \text{cm}^{-1}$ at molecular resonances of biomolecules in ultraviolet wavelengths,\cite{abdulrahman2012induced} we need a high concentrations and cm-scale pathlengths to get CD signals at the order of few millidegrees (mdeg) being within the sensitivity level of the CD spectrometers. 

Optical nanoresonators have been widely used to boost weak CD signals.\cite{garcia2013surface,warning2021nanophotonic,mohammadi2018nanophotonic,mohammadi2019accessible,mohammadi2021dual,nesterov2016role,weiss2022sensing,garcia2019enhanced,hentschel2017chiral,droulias2020absolute,raziman2019conditions,kim2022molecular}. However, nanophotonic-enhanced chiral sensing schemes operating in the visible wavelengths pose two challenges. First, the inherent chirality of matter diminishes significantly in the visible spectrum, nearly by two orders of magnitude when compared to the ultraviolet range.\cite{garcia2019enhanced} Additionally, the chiral layer interacting with the resonators has a thin profile, resulting in a very short pathlength (e.g., 10-50 nm). These two  drawbacks should be compensated by using higher concentration or higher enhancement factor due to the nanostructure. For instance, to get a CD value of 1 mdeg in the visible range with a pathlength $l = 50$ nm, we need high molar concentration $\text{C}$ = 1.6 mM and the enhancement factor $\sim 20,000$ due to the nanostructure, the latter being very challenging to reach.

One approach for nanophotonic enhanced chiral sensing is based on exposing the chiral samples to superchiral near fields generated by the resonators\cite{tang2010optical}.  In this method, the CD enhancement is proportional to the average value of the optical chirality over the volume of the chiral sample. Yet, achieving a high-average optical chirality is challenging due to its non-uniform spatial distribution\cite{mohammadi2018nanophotonic,mohammadi2019accessible}. Therefore, the optical chirality-based enhancements are typically less than two orders of magnitude. The second mechanism relies on the reverse interaction of chiral molecules toward resonators known as chirality transfer\cite{mohammadi2023nanophotonic}. The chirality transfer to achiral dielectric nanostructures can generate much stronger CD enhancement than optical chirality-based schemes. Nevertheless, achieving enhancement levels beyond 3 orders of magnitude remains challenging.

Here, we exploit thermal circular dichroism for chiral detection which is defined as $ TCD = T_R - T_L $, where $T_R$ and $T_L$ are the steady-state temperatures of the chiral sample under right- and left-circularly polarized excitations, respectively.  TCD can be understood as a two-step phenomenon: the optical component arises from the different absorbed power in chiral samples for right- and left-circularly polarized excitations, and the thermal component involves the conversion of this differential absorption into a differential temperature.  TCD combines enantiospecificity of the circular dichroism with high sensitivity of thermal measurements, achieving a superior signal-to-noise ratio for chirality detection. It has been used to detect the handedness of the chiral nanoparticles\cite{spaeth2019circular,spaeth2021photothermal,kong2018photothermal}, yet exploiting it for chiral samples remains challenging as the cross-polarizability of chiral molecules is 3-4 orders of magnitude weaker than that of nanoparticles.  We demonstrate that thermonanophotonics can significantly enhance weak TCD signals of chiral samples by leveraging opportunities in both the optical and the thermal domains.  

To gain an insight into the different factors contributing to the nanophotonic-enhanced TCD, we begin by deriving closed-form formulas for the TCD of an achiral dielectric Mie resonator covered by a thin chiral shell.  In this setup, we identify optical chirality transfer as the dominant mechanism giving rise to differential absorption inside resonators \cite{mohammadi2023nanophotonic}. Subsequently, we translate such differential absorption to the differential temperature giving rise to the TCD of the system.   Next, to further increase the TCD enhancements compared to individual resonators,  we propose arrays of resonators that are electromagnetically-uncoupled but thermally coupled. Here,  the differential absorption is taken from the solution to the individual resonators, while the thermal aspect involves solving the conduction heat equation for an ensemble of point heaters. Finally, we add electromagnetic couplings to further amplify the enhancement factors by using collective optical modes referred to as lattice resonances\cite{castellanos2019lattice}. To model such coupling  analytically, we employ Coupled Dipole Approximation (CDA) method,\cite{campione2012ewald} specifically tailored for arrays of resonators coated with chiral shells. Our proposed method, computes the differential absorption due to the chirality transfer inside each resonator within the array. Afterwards, we solve the conduction equation to find the steady state temperature profile across the array. 

 Our findings provide key insights into the interplay between optical and thermal dynamics within systems involving  chiral molecules and nanostructures. Our developed analytical methodologies allow analyzing finite chains of nanospheres covered by chiral shells which is very time-consuming to solve with numerical simulations due to their large size compared to wavelength.  We check the validity of our analytical approaches by performing Multiphysics (optical and heat) COMSOL simulations for small arrays.  

\section{Thermal Circular Dichroism for a Mie Resonator Covered by a Chiral Shell}

\begin{figure}[!t]
\centering
\includegraphics[width=\textwidth]{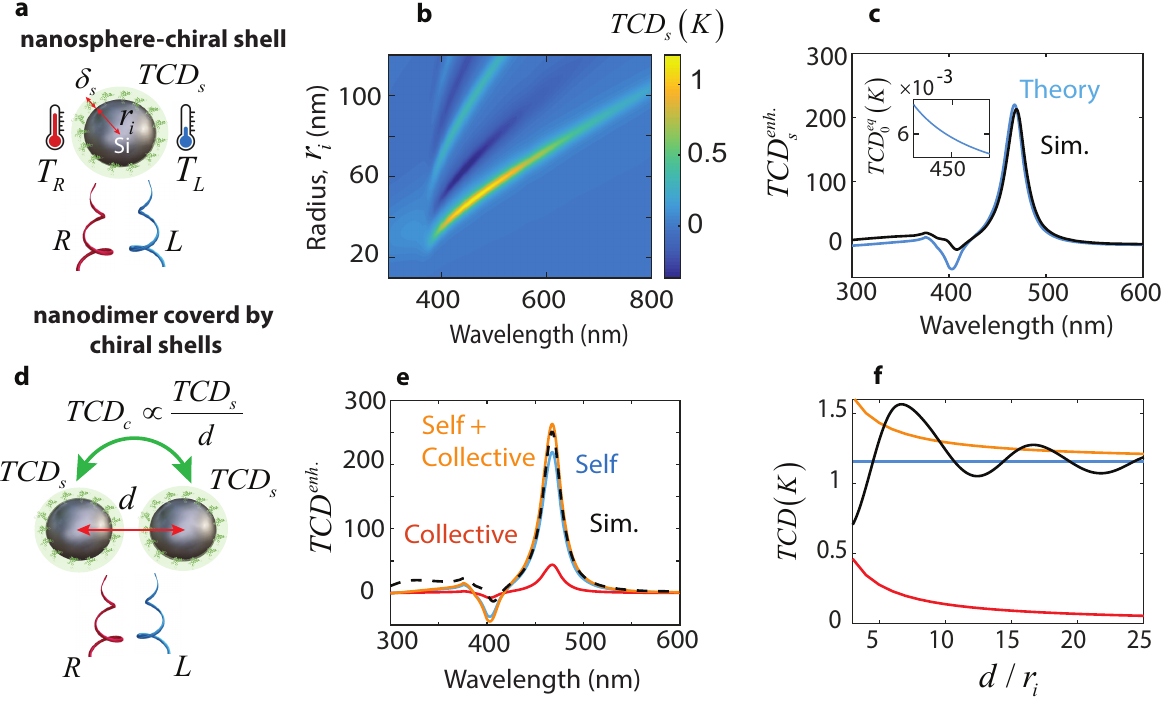}
\caption{The self- and collective-thermal circular dichroism in dielectric nanoresonators. (a) An individual silicon nanosphere of radius $r_i$ covered by a chiral shell of thickness $\delta_s$. The system is illuminated by right(R)- and left(L)-circularly polarized light of the same amplitudes, leading to different temperatures $T_R$ and $T_L$ for the sphere-shell. (b) The self-TCD of the sphere-shell (Eq.~\ref{TCD_singleNP}) as a function of wavelength and radius of the nanosphere. (c) Analytical (blue) and numerical (black) estimation of the self -TCD of the sphere-shell for $r_i = 50$ nm.   The inset shows the TCD of a chiral sphere with the same volume as the chiral shell. (d) The self- and collective-TCD in a dimer composed of two nanospheres placed at a distance $d$ (center-to-center) and coated by chiral shells. Each sphere-shell exhibits the self-TCD similar to individual resonators. There is an additional collective-TCD from the adjacent sphere-shell due to thermal conduction. (e) The different contributing parts to the TCD enhancement for one of the sphere-shells in the dimer. The blue, red, and orange curves represent the self-, collective-, and total-TCD enhancements, respectively. The black curve is obtained from numerical simulations. (f) The self-(blue), collective-(red), and total-(orange) TCD as a function of the distance between nanospheres in the dimer. The wavelength is fixed at $\lambda$=470 nm. The black curve represents the actual TCD taken from simulations.}
\label{TCD_SingleNP_and_dimer}
\end{figure}

To investigate how optical resonators can affect thermal circular dichroism, we first consider an achiral Mie resonator of radius $r_i$ covered by a chiral shell of thickness $\delta_s$ (Figure~\ref{TCD_SingleNP_and_dimer}a). The system is exposed to equally intense right($R$)- and left($L$)-circularly polarized plane waves propagating along the z-axis, described as $\textbf{E}_{R/L}=E_0\left( \hat{x} \mp i\hat{y} \right)\exp \left( -i{{k}_{0}}z \right)$, where  $k_0$ is the wavenumber. Throughout the paper, we adopt the time-harmonic convention $ e^{i\omega t}$.  The constitutive equations for a chiral medium  are $\mathbf{D}=\varepsilon \mathbf{E}-i \kappa \sqrt{\varepsilon_{0} \mu_{0}} \mathbf{H}$ and $\mathbf{B}=\mu \mathbf{H}+i \kappa \sqrt{\varepsilon_{0} \mu_{0}} \mathbf{E}$~\cite{schaferling2017chiral}, where $\kappa$ is the Pasteur parameter accounting for the coupling between the induced electric and magnetic dipoles.  We are interested in the differential temperature of the sphere-shell (i.e., TCD) which is related to the differential absorption as $\text{TCD} = \Delta P_{abs} \text{/} 4\pi K_0 r_o$, where $K_0$ is thermal conductivity of the surrounding medium, and $r_o=r_i+\delta_s$ is the outer radius of the thermal source (i.e., sphere-shell). The total differential absorption $\Delta P_{abs}$ can be decomposed into two parts: the differential absorption of the chiral shell and that of the achiral resonator. The former, is due to the optical chirality  produced by the resonator at the position of the chiral shell. The latter, being also the dominant one, is due to the chirality transfer from the chiral shell to the achiral resonator. Here, we assume that the total differential absorption equals the chirality transfer part\cite{mohammadi2023nanophotonic}. In this case, we derive the TCD as (supporting Information, Section S2):    

\begin{equation}\label{TCD_singleNP}
\begin{aligned}
  TCD &= \frac{-12 \delta_s E_0^2}{\eta_0 K_0} \left( \rho - \frac{k_0 \delta_s}{2} \right) \left( \frac{3}{2 \rho^6} + \frac{1}{\rho^4} + \frac{1}{\rho^2} \right) \\
      &\quad \times \left[ \left( \operatorname{Re}(a_1) - |a_1|^2 \right) \operatorname{Im}(\kappa b_1) + \left( \operatorname{Re}(b_1) - |b_1|^2 \right) \operatorname{Im}(\kappa a_1) \right]
\end{aligned}
\end{equation}

where $\rho ={{k}_{0}}r$, $r = r_{i}+\delta_s/2$, $\eta_0$ is the impedance of free space,  and $a_{1}$ and $b_{1}$ are the electric and magnetic dipolar Mie coefficients of the nanoparticle~\cite{ishimaru2017electromagnetic}. The subscript $"s"$ denotes the self-TCD indicating that the heat is generated by sphere-shell alone, without any influence from the collective heating effects. \cite{tsoulos2020self} We obtain the $\text{TCD}_s$ using Eq.~\ref{TCD_singleNP} as a function of radius $r_i$ and illumination wavelength (Figure~\ref{TCD_SingleNP_and_dimer}b). The Mie resonator is a silicon sphere with realistic dispersion\cite{aspnes1983dielectric} which is covered by a chiral shell of thickness $\delta_s$ = 10 nm. The system is suspended in Air with thermal conductivity $K_0 = 0.0262 \ \text{W}/\text{m} \ \text{K}$. The Pasteur parameter and the relative permittivity of chiral shell are $\kappa= (1-0.01i)\times {10}^{-2}$ and $\varepsilon_r ={1.33}^{2}-0.001i$, respectively. The incident power is $50 \ \text{MW}/\text{m}^{2}$ which corresponds  to an incident electric field $E_0 = 1.3725 \times 10^5\ \text{V/m}$. The spectra in Figure~\ref{TCD_SingleNP_and_dimer}b shows a maximized TCD of 1.2 K for radius $r_i$ = 50 nm and $\lambda$ = 470 nm which is due to the maximum chirality transfer from chiral shell to the magnetic dipolar resonance of the silicon nanosphere\cite{mohammadi2023nanophotonic}. To get the enhancement value, we normalize the TCD to that of an equivalent chiral sphere ($\text{TCD}_0^{eq}$) which has the same permittivity, Pasteur parameter, and volume as the chiral shell (Supporting Information, Section S3). By applying this normalization, we obtain the TCD enhancement for $r_i$=50 nm (Figure~\ref{TCD_SingleNP_and_dimer}c, blue curve). The inset represents the $\text{TCD}_{0}^{eq}$ reaching 0.0053 K at $\lambda$ = 470 nm. Notably, the Mie resonator amplifies this TCD by a factor of 225. To check the validity of our derivation in Eq.~\ref{TCD_singleNP}, we compute the same enhancement from numerical simulations (Figure~\ref{TCD_SingleNP_and_dimer}c, black curve), where we perform Multiphysics thermal and optical simulations. We see that the theory and simulations are in very good agreement. 
A crucial aspect of our thermonanophotonic chiral detection scheme is to ensure that the maximum temperature of the chiral molecules remains below their tolerance limit. At the same time the TCD should be above the sensitivity level of the differential thermometry measurements. The maximum temperature is the sum of the initial temperature and the temperature rise, where the latter depends linearly on the incident power. The TCD is influenced by both the incident power and the Pasteur parameter. By selecting appropriate values for these parameters, we can keep the maximum temperature within safe limits while ensuring a detectable TCD (Supporting Information, Section S4).

\section{Thermal Circular Dichroism for a Nanodimer Covered by Chiral Shells}

To obtain the enhancement beyond that achieved by single resonators, we exploit collective effects in multiple resonators. In the simplest case, we examine a nanodimer composed of two silicon spheres of radius $r_i$ = 50 nm, positioned at a center-to-center distance $d$, and both covered by chiral shells of thickness $\delta_s$ = 10 nm (Figure~\ref{TCD_SingleNP_and_dimer}d).  In this configuration, in addition to the self-TCD ($\text{TCD}_s$), there is a collective contribution ($\text{TCD}_c$) to the total-TCD of each sphere-shell. This component arises from thermal conduction which conveys the self-TCD of each sphere-shell to the neighboring particle through a factor of $1/4\pi K_0d$. We plot the self- and collective-TCD enhancements for $d=5r_i$ in Figure~\ref{TCD_SingleNP_and_dimer}e (the blue and red curves, respectively). The self part is the same as the blue curve in Figure~\ref{TCD_SingleNP_and_dimer}c, while the collective component is $\text{TCD}_c = \text{TCD}_s/4\pi K_0 d$ normalized to  $\text{TCD}_0^{eq}$.  The total enhancement (orange curve) is $(\text{TCD}_s + \text{TCD}_c)/\text{TCD}_0^{eq}$, which is aligned with the actual value taken from numerical simulations (black curve). We see that by introducing an additional particle, the TCD enhancement reaches 270, indicating a 1.2-fold increase compared to an individual sphere-shell. 

In the dimer shown in Figure~\ref{TCD_SingleNP_and_dimer}d, we overlooked electromagnetic coupling between sphere-shells. To investigate this coupling effect,  we plot the self-(blue), collective-(red), and total-(orange) TCD as a function of separation distance $d$ (Figure~\ref{TCD_SingleNP_and_dimer}f). We see that the  total-TCD can estimate the actual TCD. Yet, adding electromagnetic coupling, especially at short distances ($d<6r_i$), is crucial. For large separations, the total and actual TCDs converge due to weak electromagnetic interaction between sphere-shells. Furthermore, the total and actual TCDs are very close at $d=5r_i$, explaining  the good agreement between orange and black curves in Figure~\ref{TCD_SingleNP_and_dimer}e.

\section{Resonator Arrays: Thermally Coupled  and Electromagnetically Uncoupled}

\begin{figure}[!t]
\centering\includegraphics[width=\textwidth]{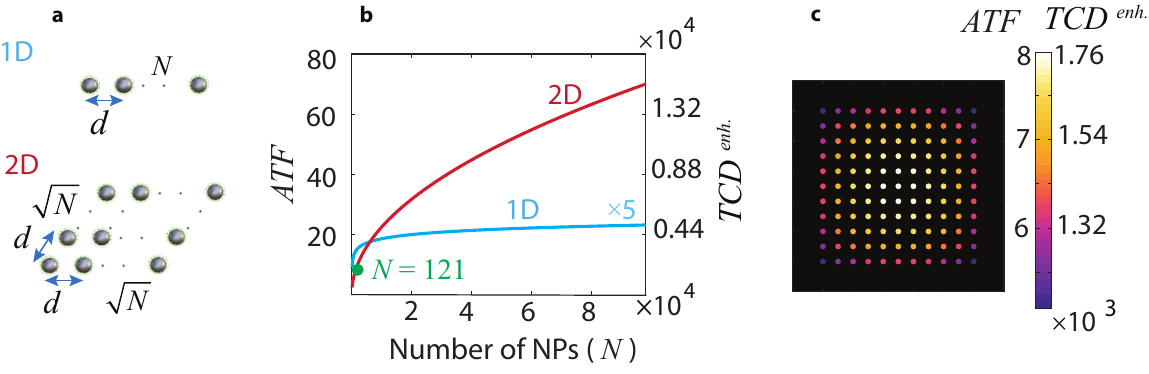}
\caption{Harnessing the collective TCD in one-(1D) and two-(2D) dimensional arrays of electromagnetically uncoupled resonators. (a) 1D and 2D arrays made of $N$ nanosphere-shells with distancing $d$ between them. (b) Array Thermal Factor (left-axis), and  $\text{TCD}^{enh}$ (right-axis) for the central sphere-shell as a function of number of nanoparticles ($N$).  Blue and red curves correspond to 1D and 2D arrays. (c) $ATF$ and $\text{TCD}^{enh}$ across the entire array for a 2D array made of $N$  = 121 (11×11) sphere-shells. For the  $\text{TCD}^{enh}$ calculations in (b) and (c), the  $\text{TCD}_s$  and  $\text{TCD}_{0}^{eq}$ are that of the individual resonator studied in Figure~\ref{TCD_SingleNP_and_dimer}c), the separation distance is  $d = 5 r_i$, and the wavelength is $\lambda = 470 nm$}
\label{TCD_arrays_noEMCoupling}
\end{figure}

The enhancement in TCD obtained through the thermal coupling, and enabled by the simple and long range  $1/d$ factor, suggests that adding more resonator can further boost this effect.\cite{naef2023light}.  If we consider an array of $N$ nanoresonators each coated by a chiral shell, then in the uncoupled electromagnetic regime, the collective-TCD for $n$-th sphere-shell is expressed as: 

\begin{equation}
TC{{D}_{c}}\left( n \right)=TC{{D}_{s}}\times \sum\limits_{i\ne n}{\frac{{{r}_{0}}}{\left\| {\textbf{d}_{i}}-{\textbf{d}_{n}} \right\|}}\text{      }i\in \left( 1:N \right),i\ne n
\end{equation}
where $\textbf{d}_i$ is position vector to  $i$-th sphere-shell and  $\text{TCD}_s$ is the self-TCD of an individual sphere-shell. Therefore, the total-TCD for $n$-th sphere-shell is  given by $TCD(n)=TCD_{s}+TCD_{c}(n)$, and the corresponding enhancement is expressed as: 

\begin{equation}\label{TCD_enh_array_noEMCoupling}
\text{ }TC{{D}^{enh}}\left( n \right)=(\frac{TC{{D}_{s}}}{TCD_{0}^{eq}})ATF\left( n \right)
\end{equation}
where $ATF$ is the Array Thermal Factor indicating the TCD enhancement factor  due to the thermal interactions within the array, and defined as: 

\begin{equation}\label{ATF_array_noEMCoupling}
ATF\left( n \right)=1+\sum\limits_{i\ne n}{\frac{{{r}_{0}}}{\left\| {\textbf{d}_{i}}-{\textbf{d}_{n}} \right\|}}
\end{equation}
 Eq.~\ref{TCD_enh_array_noEMCoupling}  breaks down the TCD enhancement into two components: the chirality transfer from the chiral shell to the underlying resonator ( $TCD_s/TCD_0^{eq}$) and the collective thermal effects ($ATF$).  As  these two components are multiplied together,  even a small increase in one can result in a substantial change in the total TCD enhancement. To demonstrate this,  we consider arrays of sphere-shells with the same parameters as in Figure~\ref{TCD_SingleNP_and_dimer} and separation distance  $d=5r_i$. We depict the $ATF$  for the central sphere-shell in 1D (blue curve) and 2D (red curve) arrangements as a function of the number of nanoparticles $N$ (left axis in Figure~\ref{TCD_arrays_noEMCoupling}b). We see that by increasing the number of sphere-shells to $10^4$ the $ATF$ for 1D and 2D arrays can reach values up to 4.6 and 70, respectively. Considering an operating wavelength of $\lambda$ = 470 nm, multiplying this $ATF$ by the TCD enhancement of an individual sphere-shell (i.e., 220), yields the corresponding TCD enhancements within the array as 1018 and 1.54 $\times 10^4$ , respectively (right axis in Figure~\ref{TCD_arrays_noEMCoupling}b). To investigate the position dependency in the array, we depict  the  $ATF$ and $TCD^{enh}$ across a 2D array composed of 121 nanoparticles arranged in an $11\times11$ lattice with a spacing  $d=5r_i$ (Figure~\ref{TCD_arrays_noEMCoupling}c). We see that the maximum enhancement occurs at the center of the array which is always the case when we have no electromagnetic coupling between resonators. Furthermore, in the center of the array  $ATF \approx$ 8 and $TCD^{enh}\approx $ 1756 corresponding to the green dot in Figure~\ref{TCD_arrays_noEMCoupling}b).

\section{Resonator Arrays: Thermally and Electromagnetically Coupled}

The analytical prediction of  TCD in the uncoupled electromagnetic regime starts deviating from numerical simulation for $d < 6 r_i$  (Figure~\ref{TCD_SingleNP_and_dimer}f).  On the other hand,  $1/d$ decay of the collective-TCD suggests to keep separation distances small to effectively capture thermal interactions. This underscores the importance of considering electromagnetic couplings to make more accurate analytical approximations for shorter separation distances. 

To accomplish this, we must characterize the differential absorption of each sphere-shell within the array. This serves as the source term for the thermal conduction equation, enabling us to compute the self-, collective-, and total-TCD of each sphere-shell. We begin our approach by applying Couple Dipole Approximation (CDA) method to the array of resonators without the chiral shells (Supporting Information, Section S5). In the CDA method, we replace each sphere by an electric and a magnetic dipole and then the electromagnetic interaction between these dipoles is computed through solving a self-consistent system of equations. The output of this system is the local fields ($\boldsymbol{E}_{loc,R/L}^{w.o}$ ,  $\boldsymbol{H}_{loc,R/L}^{w.o}$) at the position of each sphere and the corresponding induced dipole moments ($\boldsymbol{p}_{R/L}^{w.o}$ ,  $\boldsymbol{m}_{R/L}^{w.o}$) . The superscript "$w.o$" denotes that the quantities are associated to the sphere array alone and excluding the chiral shells.  As the  sphere array alone is achiral, such local fields and dipole moments have the same amplitudes for right- and left-circular polarizations (i.e.,   $|\boldsymbol{E}_{loc,R}^{w.o}|=|\boldsymbol{E}_{loc,L}^{w.o}|$, $|\boldsymbol{H}_{loc,R}^{w.o}|=|\boldsymbol{H}_{loc,L}^{w.o}|$,$|\boldsymbol{p}_{R}^{w.o}|=|\boldsymbol{p}_{L}^{w.o}|$,$|\boldsymbol{m}_{R}^{w.o}|=|\boldsymbol{m}_{L}^{w.o}|$). Using the free-space Green's functions and the induced dipole moments $\boldsymbol{p}_{R/L}^{w.o}$ and $\boldsymbol{m}_{R/L}^{w.o}$, we can calculate the near field around each resonator ($\boldsymbol{E}_{s,R/L}$ and $\boldsymbol{H}_{s,R/L}$ ) and at the position of the surrounding  chiral shell. Next, we cover the resonators with chiral shells and assume that the presence of the shell does not disturb   $\boldsymbol{E}_{s,R/L}$ and $\boldsymbol{H}_{s,R/L}$ .  This assumption is correct as far as the chiral shells are thin and their refractive index is close to the surrounding medium.  Given this assumption, the electric ($\boldsymbol{P}_{s,R/L}$) and magnetic ($\boldsymbol{M}_{s,R/L}$) polarization densities  over the chiral shells are related to the near fields as $\boldsymbol{P}_{s,R/L}=\epsilon_0(\epsilon_r-1)\boldsymbol{E}_{s,R/L} - i\kappa\boldsymbol{H}_{s,R/L}/c_0$ and $\boldsymbol{M}_{s,R/L}= i\kappa\boldsymbol{E}_{s,R/L}/\eta_0$, respectively.  To calculate the chirality transfer, we need to find the back-action of such polarization on the underlying resonator. This results in the new local fields and the dipole moments  for right- and left-circular excitations being different now due to the presence of chirality in the system  (i.e.,   $|\boldsymbol{E}_{loc,R}^{w}|\neq|\boldsymbol{E}_{loc,L}^{w}|$, $|\boldsymbol{H}_{loc,R}^{w}|\neq|\boldsymbol{H}_{loc,L}^{w}|$,$|\boldsymbol{p}_{R}^{w}|\neq|\boldsymbol{p}_{L}^{w}|$,$|\boldsymbol{m}_{R}^{w}|\neq|\boldsymbol{m}_{L}^{w}|$). The superscript "$w$" indicates that the quantities correspond to the sphere array coated with chiral shells. We can obtain the extinct and scattered powers by each sphere-shell as: 
 \begin{equation}
    P_{R/L}^{ext} = \frac{\omega_0}{2} \operatorname{Im} (\boldsymbol{E}_{loc,R/L}^{w.o}.\boldsymbol{p}_{R/L}^{w} + \mu_0 \boldsymbol{H}_{loc,R/L}^{w.o}.\boldsymbol{m}_{R/L}^{w})
\end{equation}
\begin{equation}
    P_{R/L}^{sca} = \frac{\eta_0 k_0^4}{12\pi} (c_0^2|\boldsymbol{p}_{R/L}^{w}|^2 + |\boldsymbol{m}_{R/L}^{w}|^2)
\end{equation}

where $\omega_0$ is the angular frequency.  Lastly, the differential extinct, scattered, and absorbed powers due to chirality transfer can be expressed as $\Delta P^{ext} = P_{R}^{ext}-P_{L}^{ext}$, $\Delta P^{sca} = P_{R}^{sca}-P_{L}^{sca}$, and $\Delta P^{abs} = \Delta P^{ext} - \Delta P^{sca}$, where the latter is of particular interest for our problem as it determines the self-TCD of $n$'th sphere-shell as:  

\begin{equation}
    TCD_s (n) = \frac{\Delta P_{abs}(n)}{4 \pi K_0 r_o}
\end{equation}
Then, the total TCD for the $n$-th sphere-shell reads as: 
\begin{equation}
TCD(n)=TCD_s (n) +  \sum\limits_{i\ne n}{\frac{TCD_s (i)}{||d_i-d_n||/r_o}} \text{      }     \text{       }   i\in \left( 1:N \right),i\ne n
\end{equation}

\begin{figure}[t!]
\centering\includegraphics[width=6in]{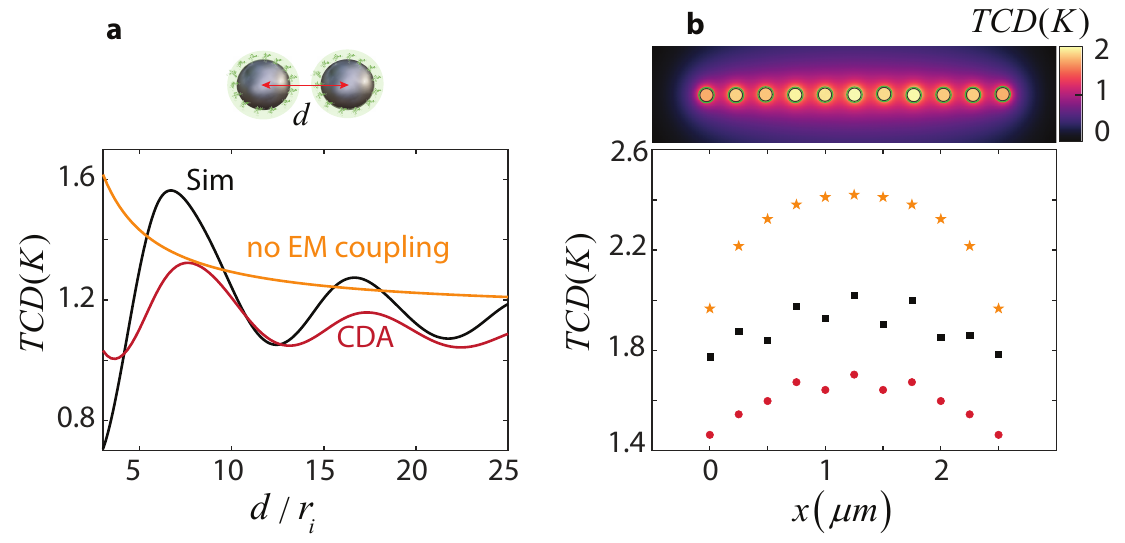}
\caption{The effect of electromagnetic coupling on the TCD of sphere-shell array.  (a) The TCD of one of the sphere-shells for the dimer  in Figure~\ref{TCD_SingleNP_and_dimer}d, as a function of separation distance $d$. The orange curve corresponds to the case with no electromagnetic coupling (same as orange curve in Figure~\ref{TCD_SingleNP_and_dimer}f). The Red curve is obtained from CDA, and the black curve is actual TCD based on numerical simulations (same as black curve in Figure~\ref{TCD_SingleNP_and_dimer}f). (b) The TCD across a 1D array made of  $N = 11$ sphere-shells separated by distance $d = 5r_i$. The parameters of the system are the same as Figure~\ref{TCD_SingleNP_and_dimer}) and the operating wavelength is $\lambda = 470nm$. Top panel shows the simulated TCD. In the bottom panel, the orange stars indicate TCD in the uncoupled electromagnetic regime, the red circles correspond to the TCD obtained from CDA, and the black rectangles denote the TCD obtained from numerical simulations.}
\label{TCD_EMcouling}
\end{figure}

To check the validity of the proposed method, we examine the same dimer in Figure~\ref{TCD_SingleNP_and_dimer}d and obtain the TCD of one of the nanoparticles as a function of separation distance $d$  (red curve in Figure~\ref{TCD_EMcouling}a) . Comparing the result from CDA with that obtained based on no electromagnetic coupling (orange curve), we observe that the former aligns more closely with the actual TCD (black curve), thereby offering more accurate predictions, particularly for small separation distances. Next, we investigate the TCD profile across a 1D array made of $N=11$ sphere-shells placed at distance $d=5r_i$ (top panel in Figure~\ref{TCD_EMcouling}b). The parameters for spheres and shells are the same as previous sections, and the wavelength is $\lambda$ = 470 nm. Again, we obtain the TCD using CDA method (red circles in bottom panel in Figure~\ref{TCD_EMcouling}b) and compare it to the case when there is no electromagnetic coupling in the array (orange stars) as well as to the actual TCD taken from numerical simulations (black rectangles). Notably, we see that the CDA  method not only aligns closer to the simulation but also captures the real temperature profile. 

We should note that there is yet a discrepancy between CDA-based approximations and numerical simulations. This error could be attributed to the differential absorption inside the chiral shells due to the optical chirality, which is neglected because of the dominance of chirality transfer-based differential absorption. Furthermore, the contrast in the refractive index of the chiral sample ($1.33^2 - 0.001i$ ) and the surrounding medium (in this case, air), contributes to the error.  This alters the actual polarizing near fields acting on the chiral shells compared to those of the array without chiral shells (i.e., $\boldsymbol{E}_{s,R/L}$ and $\boldsymbol{H}_{s,R/L}$). 

\section{Combining Collective Thermal Interactions and Optical Lattice Resonances}

\begin{figure}[t!]
\centering\includegraphics[width=6 in]{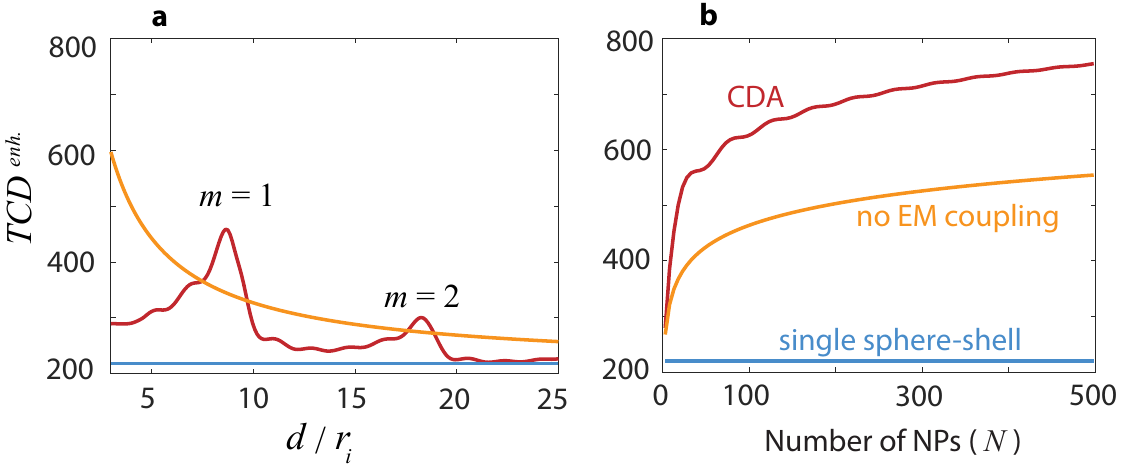}
\caption{Synergy of collective thermal interactions and optical lattice resonances in a 1D array of sphere-shells. The TCD enhancement in the central sphere-shell, (a) as a function of separation distance $d$ for $N=11$ nanoparticles, and (b) as a function of number of nanoparticles ($N$) for $d=8.6r_i$. In both (a) and (b) the red curve corresponds to the CDA-based calculation. The orange curve shows enhancement without electromagnetic coupling. The blue curve denote the base-enhancement  of an individual sphere-shell.}
\label{LRs}
\end{figure}

In Figure~\ref{TCD_EMcouling}b, we noted that the TCD achieved by the 1D array made of $N$ = 11 sphere-shells with a spacing $d=5r_i$ reaches 1.7, corresponding to 320-fold enhancement. To further boost electromagnetic interactions,  we propose lattice resonances (LRs) emerging from the radiative coupling between nanoparticles through in-plane diffraction\cite{murai2020enhanced}. For an infinite array the $m$-th order of such collective Mie modes occurs at spacing $d = m\lambda$. To verify this, we calculate the TCD enhancement using CDA as a function of spacing in the central sphere-shell of the array examined in Figure~\ref{TCD_EMcouling}b  (red curve in Figure~\ref{LRs}a). The first and the second order LRs are evident for  $d (m=1)=8.6r_i$ and $d(m=2)=18.25r_i$, corresponding to  $d=0.91\lambda$ and $d=1.94\lambda$, as expected. For comparison, we also depict the TCD enhancement in the uncoupled electromagnetic regime (orange curve) as well as  the baseline enhancement provided by an individual resonator (blue curve). We see that the LRs can increase the TCD beyond the uncoupled regime reaching 500-fold enhancement at the strongest order (i.e., $m=1$).   Next, we investigate the effect of number of nanoparticles ($N$) on the TCD for a fixed spacing  $d (m=1)=8.6r_i$ (Figure~\ref{LRs}b). We see that by increasing $N$ higher enhancement factors up to 700-fold are possible.  We anticipate that for the 2D arrays of sphere-shells, LRs could achieve TCD enhancement factors even greater than the corresponding non-coupled regime (see red curve in Figure~\ref{TCD_arrays_noEMCoupling}b ).

 In conclusion, we suggested thermal circular dichroism combined with nanophotonics, for ultrasensitive detection of molecular chirality. We demonstrated our thermonanophotonic approach analytically and numerically for very thin chiral coatings on spherical silicon resonators. We identified the chirality transfer as the main underlying source of TCD, and based on this,  we derived closed-form formulas for the TCD of an individual resonator-shell.  Furthermore,  in our quest towards higher enhancement factors, we proposed arrays, where the thermal and optical properties of the problem can be synergistically combined.  We showed that the collective thermal effects in combination with the optical lattice resonance can increase the TCD substantially beyond the individual resonator level. Our findings open new avenues for advancements in ultrasensitive chiral detection.

\begin{acknowledgement}


\end{acknowledgement}

\begin{suppinfo}
 
\end{suppinfo}

\bibliography{achemso-demo}

\end{document}